\documentclass[journal=ancac3,manuscript=article]{achemso}

\usepackage{chemformula} 
\usepackage[T1]{fontenc} 
\usepackage{subcaption}
\usepackage{xr-hyper}
\title{Atomic-Scale Origins of Oxidation Resistance in Amorphous Boron Nitride}

\author{Onurcan Kaya}
\affiliation[ICN2]{Catalan Institute of Nanoscience and Nanotechnology (ICN2), CSIC and BIST, Campus UAB, 08193 Bellaterra, Barcelona, Spain}
\alsoaffiliation[RMIT]{School of Engineering, RMIT University, Melbourne, Victoria 3001, Australia}
\alsoaffiliation[UAB]{Department of Electronic Engineering, Universitat Autònoma de Barcelona (UAB), Campus UAB, 08193 Bellaterra, Barcelona, Spain}
\author{Qiushi Deng}
\affiliation[ANU]{School of Engineering, The Australian National University, Canberra, ACT 2600, Australia}
\author{Thomas Souvignet}
\affiliation[UCBL]{Université Claude Bernard Lyon 1, CNRS, LMI UMR 5615, 69100 Villeurbanne, France}
\author{Catherine Marichy}
\affiliation[UCBL]{Université Claude Bernard Lyon 1, CNRS, LMI UMR 5615, 69100 Villeurbanne, France}
\author{Catherine Journet}
\affiliation[UCBL]{Université Claude Bernard Lyon 1, CNRS, LMI UMR 5615, 69100 Villeurbanne, France}
\author{Ivan Cole}
\affiliation[ANU]{School of Engineering, The Australian National University, Canberra, ACT 2600, Australia}
\author{Stephan Roche}
\affiliation[ICN2]{Catalan Institute of Nanoscience and Nanotechnology (ICN2), CSIC and BIST, Campus UAB, 08193 Bellaterra, Barcelona, Spain}
\alsoaffiliation[ICREA]{ICREA, 08010 Barcelona, Spain}
\email{stephan.roche@icn2.cat}


\begin{document}

\begin{tocentry}

\centering
\includegraphics[width=\linewidth]{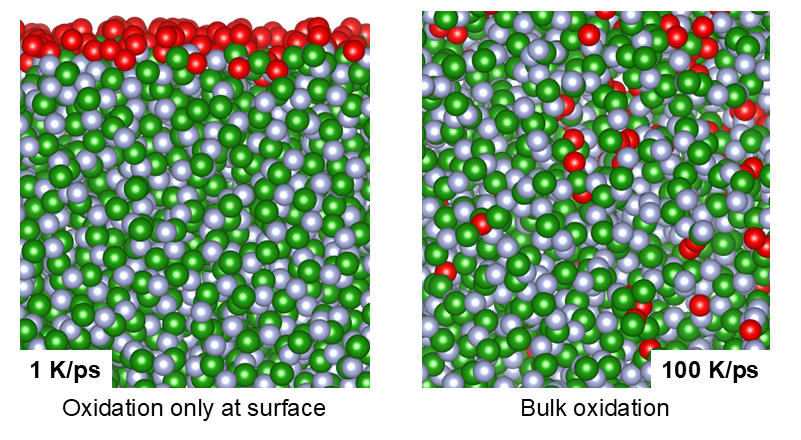}

\end{tocentry}

\begin{abstract}
Amorphous boron nitride (\textrm{$\alpha$}-BN) is a promising ultrathin barrier for nanoelectronics, yet the atomistic mechanisms governing its chemical stability remain poorly understood. Here, we investigate the structure-property relationship that dictates the oxidation of \textrm{$\alpha$}-BN using a combination of machine-learning molecular dynamics simulations and angle-resolved X-ray photoelectron spectroscopy. The simulations reveal that the film structure, controlled by synthesis conditions, is the critical factor determining oxidation resistance. Dense, chemically ordered networks with a high fraction of B-N bonds effectively resist oxidation by confining it to the surface, whereas porous, defect-rich structures with abundant homonuclear B-B and N-N bonds permit oxygen penetration and undergo extensive bulk degradation. These computational findings are consistent with experimental trends observed in \textrm{$\alpha$}-BN films grown by chemical vapour deposition. XPS analysis shows that a film grown at a higher temperature develops a more ordered structure with a B/N ratio nearer to stoichiometric and exhibits superior resistance to surface oxidation compared to its more defective, lower-temperature counterpart. Together, these results demonstrate that the oxidation resistance of \textrm{$\alpha$}-BN is a tunable property directly linked to its atomic-scale morphology, providing a clear framework for engineering chemically robust dielectric barriers for future nanoelectronic applications.
\end{abstract}

\section{Introduction}

In advanced nanoelectronic devices, chemical degradation of materials, particularly oxidation, has become a critical reliability concern. As components shrink to the nanometer scale, their high surface-to-volume ratio makes them more vulnerable to oxygen exposure, leading to poor performance and failure over time\cite{Kaloyeros2000, Hong2020}. Conventional dielectrics and barrier layers (such as silicon oxides or nitrides) are reaching fundamental limits in the ultrathin regime. At reduced dimensions, insufficient mechanical robustness and chemical instability promotes high interface defect density and rapid oxidation, which ultimately causing device failure \cite{sattari2025, Lo2018beol, chen2023tailoring}. This has prompted an intense search for new materials that can serve as ultrathin passive and encapsulation layers, protecting sensitive nanoelectronic structures from the ambient environment. \\
Boron nitride in its hexagonal crystalline phase (h-BN) is chemically robust, with atomically thin films stable in air up to $   1100,^\circ C $ under reported conditions \cite{Li2014, kaya2025literature, Lo2017, Dmitri2010}. This performance together with a wide band gap of $   6\ eV $ and a chemically stable surface makes hBN attractive as a two dimensional barrier material \cite{Cassabois2016, Glavin2014}. Large scale use remains challenging because growth typically needs high temperature, occurs on specific substrates with limited areas, and grain boundaries or defects can compromise protection \cite{Hong2020}. In contrast, amorphous boron nitride (\textrm{$\alpha$}-BN) retains many of the desirable properties of hBN while offering improved processability\cite{Hong2020}. It can be deposited uniformly at low temperatures (65 ºC by atomic layer deposition (ALD) \cite{chen2023tailoring}, $\sim$ 250 ºC by chemical vapour deposition (CVD) \cite{sattari2025}) with wafer-scale conformality and no grain boundaries\cite{Hong2020, Glavin2014}. Because low heat input is required for \textrm{$\alpha$}-BN growth, it can be grown on many substrates, including two-dimensional materials and polymers\cite{Glavin2016}. These traits make \textrm{$\alpha$}-BN compatible with standard silicon process flows and CMOS integration. Experimental studies report that ultrathin \textrm{$\alpha$}-BN films a few nanometers thick show low dielectric constant $   k<2 $, good electrical integrity with minimal leakage, and strong resistance to metal diffusion \cite{Hong2020, chen2023tailoring, Kim2023, kaya2024amorphous}. At comparable thickness these properties match or exceed those of conventional $   SiO_2 $ insulators \cite{Akkili2025, Hong2020}. \textrm{$\alpha$}-BN is also effective as a capping layer for air sensitive two dimensional materials. Encapsulation of graphene with \textrm{$\alpha$}-BN preserves structural and electronic quality under ambient humidity and improves carrier mobility and long term stability relative to unprotected devices \cite{SattariEsfahlan2023, sattari2025}. Similar behavior is reported for other two dimensional systems after prolonged air exposure \cite{chen2023tailoring}. These findings establish \textrm{$\alpha$}-BN as a promising ultrathin barrier, yet the atomistic mechanisms that govern its oxidation resistance remain poorly understood.\\
 Although \textrm{$\alpha$}-BN films have been used as encapsulation layers\cite{SattariEsfahlan2023, chen2023tailoring} and passivation layers\cite{Lu2022lowtemp} in nanoelectronic devices and exhibit protective behavior, the atomistic factors that govern this behavior remain unclear. Oxidation kinetics of boron nitride compounds are sensitive to crystallinity, porosity and defect content, which influence oxygen uptake and bonding pathways\cite{LAVRENKO198625, Taniguchi2003, Ismach2012}. In amorphous BN, this sensitivity is expected to be greater since its disordered network can contain diverse local coordination (\textrm{$sp$}, \textrm{$sp^2$}, \textrm{$sp^3$}), homonuclear bonds, and deviations from ideal B/N ratio within films depending on growth parameters\cite{sattari2025, cheng1987}. Some regions may contain under-coordinated sites or excess boron that promote oxygen uptake. Recent experiments indicate that nitrogen vacancies create locally boron-rich environments, where oxygen incorporates and oxidizes the excess boron \cite{chen2023tailoring, lu2023one, Miao2024}. Device studies on amorphous BN memristors also identify boron vacancies as active centres, reinforcing vacancy-mediated reactivity under oxidizing conditions\cite{Khot2022,Lee2020synaptic}. In addition, films prepared by plasma-enhanced CVD can incorporate hydrogen in the form of B–H and N–H bonds; these species are metastable and can break during annealing, creating dangling bonds that act as reactive sites for oxygen uptake\cite{ackerman1995}. Infrared and electron spin resonance studies further indicate that hydrogen evolution leads to defect formation and structural rearrangement, which accelerates degradation under ambient exposure\cite{lin1995, brown1997}. As a result, residual hydrogen compromises the stability of \textrm{$\alpha$}-BN and provides additional pathways for oxidative attack. Oxidation in \textrm{$\alpha$}-BN is governed by the local distribution of reactive sites, including vacancies, coordination defects, residual B–H/N–H species and regions with local strain. Direct observation of these events is difficult, because oxygen uptake is highly local and evolves on short timescales that most \textit{in situ} methods cannot resolve.\\
Direct observation of oxidation processes in amorphous materials remains challenging because oxygen uptake and structural changes occur locally and evolve rapidly, beyond the resolution of \textit{in situ} methods\cite{Villena2024}. We therefore use atomistic simulation to investigate the atomistic mechanisms of oxidation. However, conventional molecular dynamics based on empirical force fields often lacks the flexibility to describe bond breaking and formation across diverse local environments, limiting applicability to reactive processes in disordered systems \cite{Behler2016,Deringer2017}. First principles methods such as density functional theory (DFT) provide accurate treatment of chemical bonding but are computationally constrained in both time and length scales\cite{Deringer2017}. Machine-learning interatomic potentials including the Gaussian Approximation Potential (GAP)\cite{Bartok2010,Bartok2013,Bartok2015} enable reactive atomistic simulations at larger length and timescales than first-principles methods, while capturing bond breaking and formation in disordered networks\cite{Behler2016,Deringer2017}. Recent studies show that such models can approach DFT accuracy for relevant configurations and extend simulations to longer time scales \cite{Batzner_2022,NEURIPS2022_4a36c3c5,batatia2025foundationmodelatomisticmaterials}. \\

\section{Results and Discussion}
We investigate how morphological features in \textrm{$\alpha$}-BN influence its interaction with oxygen using machine-learning molecular dynamics simulations complemented by experimental characterization. Simulations based on a machine-learned GAP potential quantify how coordination defects, homonuclear bonding, mass density, and the B/N ratio affect oxygen uptake and the associated structural changes. In parallel, angle-resolved XPS on CVD-grown \textrm{$\alpha$}-BN films with near-stoichiometric B/N ratio composition tracks changes in B 1s and N 1s spectra at multiple emission angles following ambient exposure. Both approaches reveal consistent behavior. Dense, well-connected B–N networks confine oxidation near the surface, whereas films with greater defect content and porosity permit deeper oxygen penetration and more extensive structural modification. The sections that follow first establish how growth conditions affect the morphology of \textrm{$\alpha$}-BN, and then analyze the oxidation response through simulation and experiment.

\subsection{Morphology of Amorphous BN Films}  \label{s:morphology}

\begin{figure}[htbp]
  \centering
  \begin{subfigure}[t]{0.45\textwidth}
    \centering
    \includegraphics[width=\linewidth]{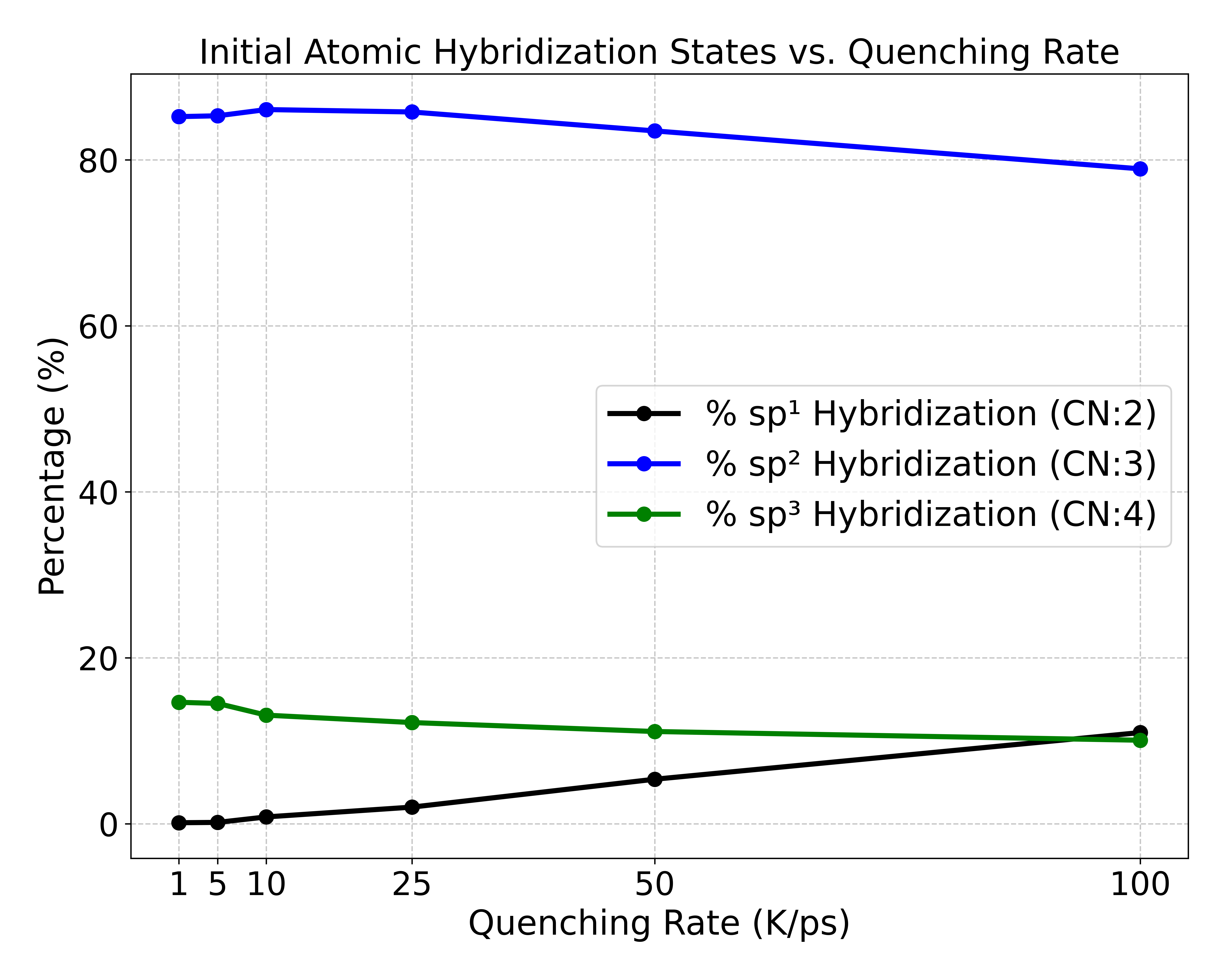}
    \caption{}
    \label{fig:hybridmorp} 
  \end{subfigure}
  \hspace{0.05\textwidth}
  \begin{subfigure}[t]{0.45\textwidth}
    \centering
    \includegraphics[width=\linewidth]{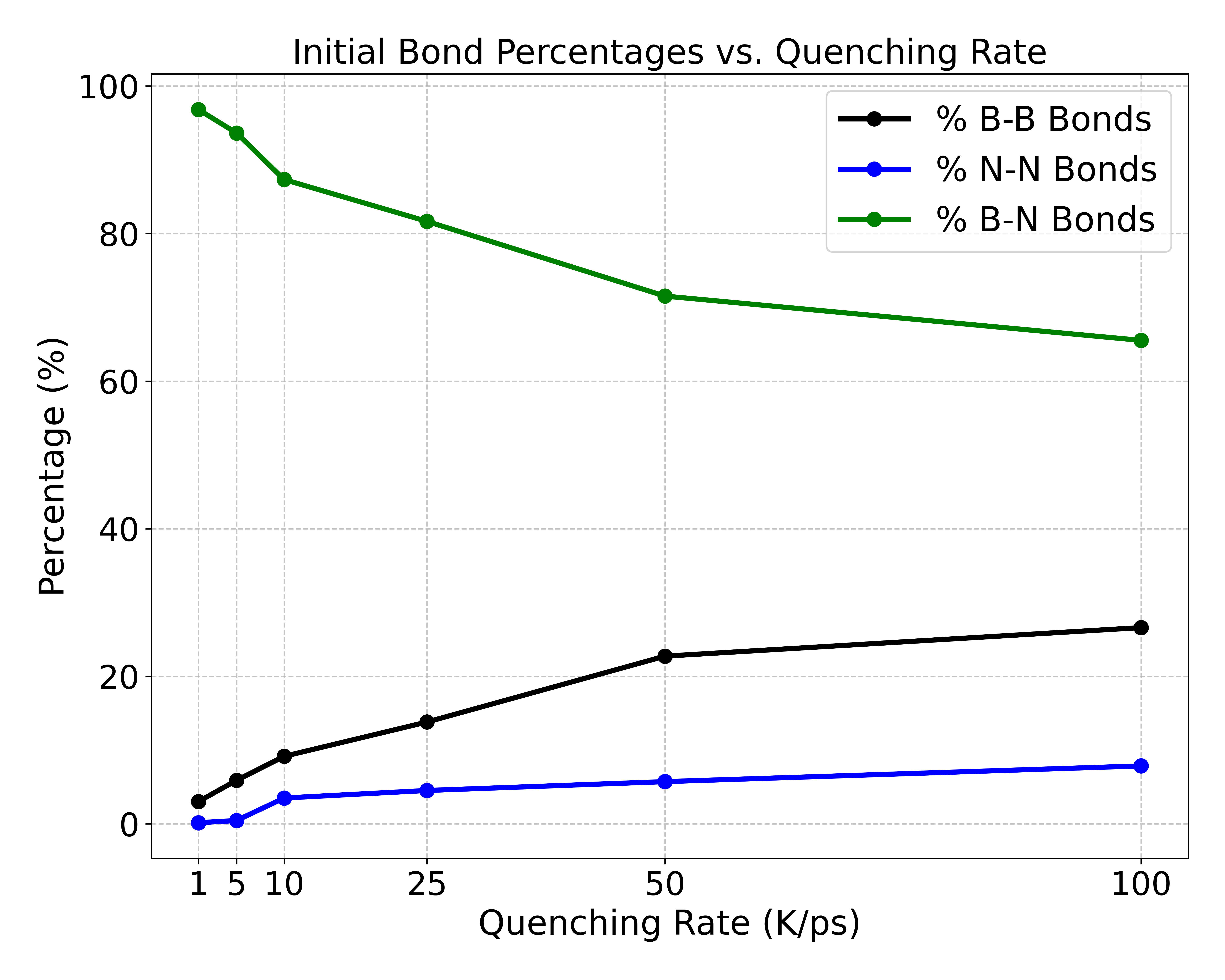}
    \caption{}
    \label{fig:bondingsmorp}
  \end{subfigure}
  \vspace{1em}
  \begin{subfigure}[t]{0.45\textwidth}
    \centering
    \includegraphics[width=\linewidth]{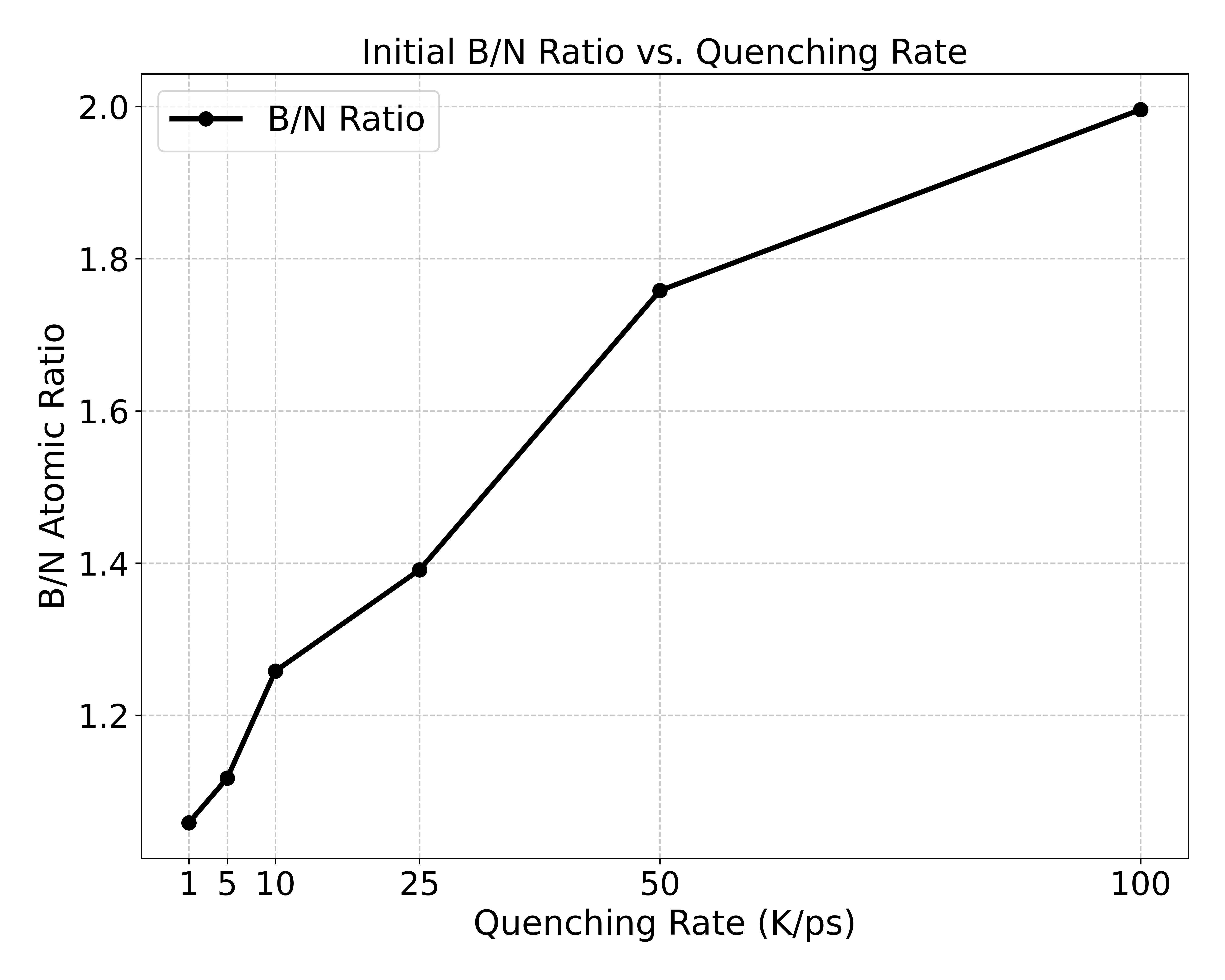}
    \caption{}
    \label{fig:BNmorp} 
  \end{subfigure}
  \hspace{0.05\textwidth}
  \begin{subfigure}[t]{0.45\textwidth}
    \centering
    \includegraphics[width=\linewidth]{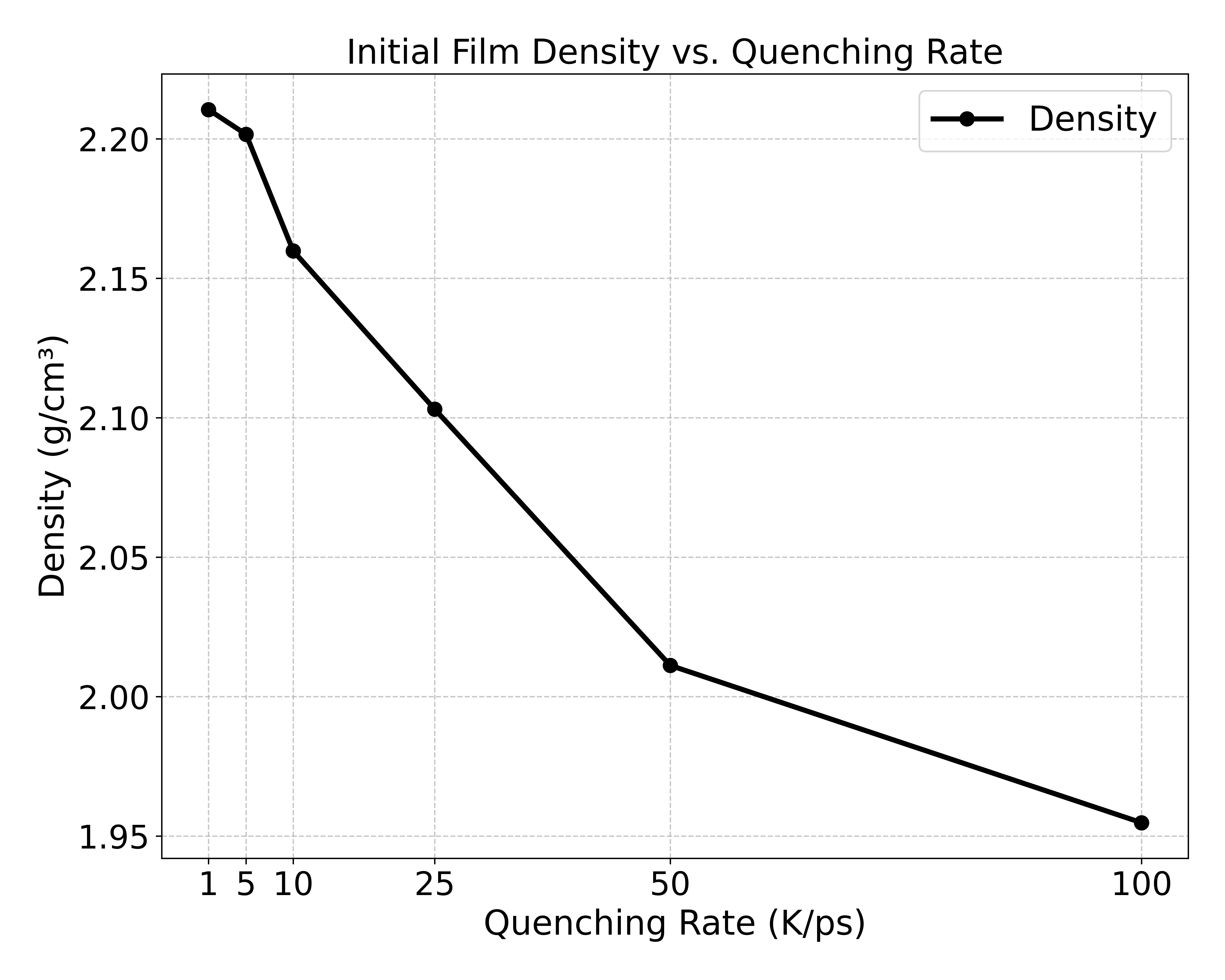}
    \caption{}
    \label{fig:densitymorph} 
  \end{subfigure}
  \caption{ Impact of quenching rate on the initial structure of \textrm{$\alpha$}-BN films. (a) Fraction of atoms in \textrm{$sp$}, \textrm{$sp^2$}, and \textrm{$sp^3$} hybridization states. (b) Percentage of B–N, B–B, and N–N bonds. (c) B/N atomic ratio. (d) Mass density (g cm$^{-3}$).}
  \label{fig:morphology_vs_quench}
\end{figure}

The quenching rate (QR) during melt–quench synthesis governs the atomic arrangement, bonding statistics, and morphological stability of \textrm{$\alpha$}-BN films. Varying QR from 1 to 100 K ps$^{-1} $ produces systematic trends in four structural descriptors. These include the hybridization state (\textrm{$sp$}, \textrm{$sp^2$}, \textrm{$sp^3$}), the distribution of B–N, B–B, and N–N bonds, the B/N atomic ratio, and the mass density. Faster cooling results in more under-coordinated sites, increased homonuclear bonding, and reduced density, whereas slower cooling promotes the formation of a chemically ordered heteronuclear B–N network. These trends are shown in Figure \ref{fig:morphology_vs_quench}\subref{fig:hybridmorp}-\subref{fig:densitymorph}. Reported values are averages over five independent trajectories, and error bars indicate standard deviations. Bonding and coordination are defined using a cutoff radius of $ r_{c} = 1.90\ \textrm{\AA}$. \\
The structural features of \textrm{$\alpha$}-BN films evolve systematically with quenching rate. At the lowest rates (1–5 K ps$^{-1}$ ), the system has sufficient time to relax during cooling, yielding chemically ordered, low-energy configurations. The films are predominantly \textrm{$sp^2$}-coordinated (exceeding 85\%), with a minor \textrm{$sp^3$} fraction ($\sim$14.5\%) that provides local three-dimensional cross-linking (Figure~\ref{fig:hybridmorp}); under-coordinated \textrm{$sp$} sites are nearly absent ($<$0.2\%). More than 93\% of all bonds are heteronuclear B–N, while homonuclear B–B and N–N bonds remain minimal (Figure~\ref{fig:bondingsmorp}), consistent with near-stoichiometric and chemically homogeneous films (Figure~\ref{fig:BNmorp}). Taken together, these features indicate a relaxed amorphous state that is consistent with a continuous random network. As the QR increases to intermediate values (10–25 K ps$^{-1}$) the time available for structural relaxation decreases and the structure becomes less relaxed than in the slow-quenched case. The \textrm{$sp^2$} fraction declines and under-coordinated \textrm{$sp$} sites become more prevalent as shown in Figure~\ref{fig:hybridmorp}. Faster quenching limits local rearrangement and diffusion of atoms promoting the formation of B-B and N-N clusters as in Figure~\ref{fig:bondingsmorp}. These deviations from more stable configurations become more prominent at even higher QR (50-100 K ps$^{-1}$). The \textrm{$sp$} fraction reaches about 11\%. B–N bonding falls below 66\%, so that roughly one third of bonds are homonuclear B–B and N–N (Figures~\ref{fig:hybridmorp} and \ref{fig:bondingsmorp}). At these rates, the structures behave as frozen liquid, a metastable configuration that is trapped at higher energy. These structural changes also alter stoichiometry. While slow-quenched films have a stoichometric like structure (B/N ratio of $<$1.10), this increases to the 1.2–1.4 range for intermediate values and approaches ~2.0 when QR reaches 100 K ps$^{-1}$. Since N-N bonds are less stable, they are usually broken during the annealing step leading to B-rich films. This nitrogen loss arises primarily from the instability and rupture of N–N bonds during annealing. B-B clusters, due to their relatively strong covalent character and high bond energy, tend to persist, whereas N–N bonds tend to break more readily, leading to N depletion. Thus, the QR does not merely control structural and chemical disorders, but directly dictates elemental composition through dynamic selection of bond stability during the quench. We also observe a steady reduction in mass density, which decreases from approximately 2.21 g cm$^{-3}$ to 1.95 g cm$^{-3}$ as shown in Figure \ref{fig:densitymorph}. This suggests a growth in free volume, structural voids, or low-density domains. The drop in density, coupled with the rise in under-coordinated atoms, strongly implies increased nanoporosity and open network regions. These low-density regions are not uniformly distributed. They typically emerge near surfaces or in areas rich in homonuclear clusters, further compromising film stability.\\
These findings collectively indicate that the quench rate governs the morphology of the amorphous network. As the quench rate increases, the number of under‑coordinated sites rises, homonuclear B–B and N–N bonds become more prevalent, the composition shifts towards B‑rich, and the density decreases, all of which is consistent with increased free volume and emerging nanoporosity. These features reduce structural and chemical stability and open pathways for oxygen incorporation under oxidative conditions.

\subsection{Oxidation Behavior of Amorphous BN Films }

\begin{figure}[htbp]
\centering
\includegraphics[width=0.40\columnwidth]{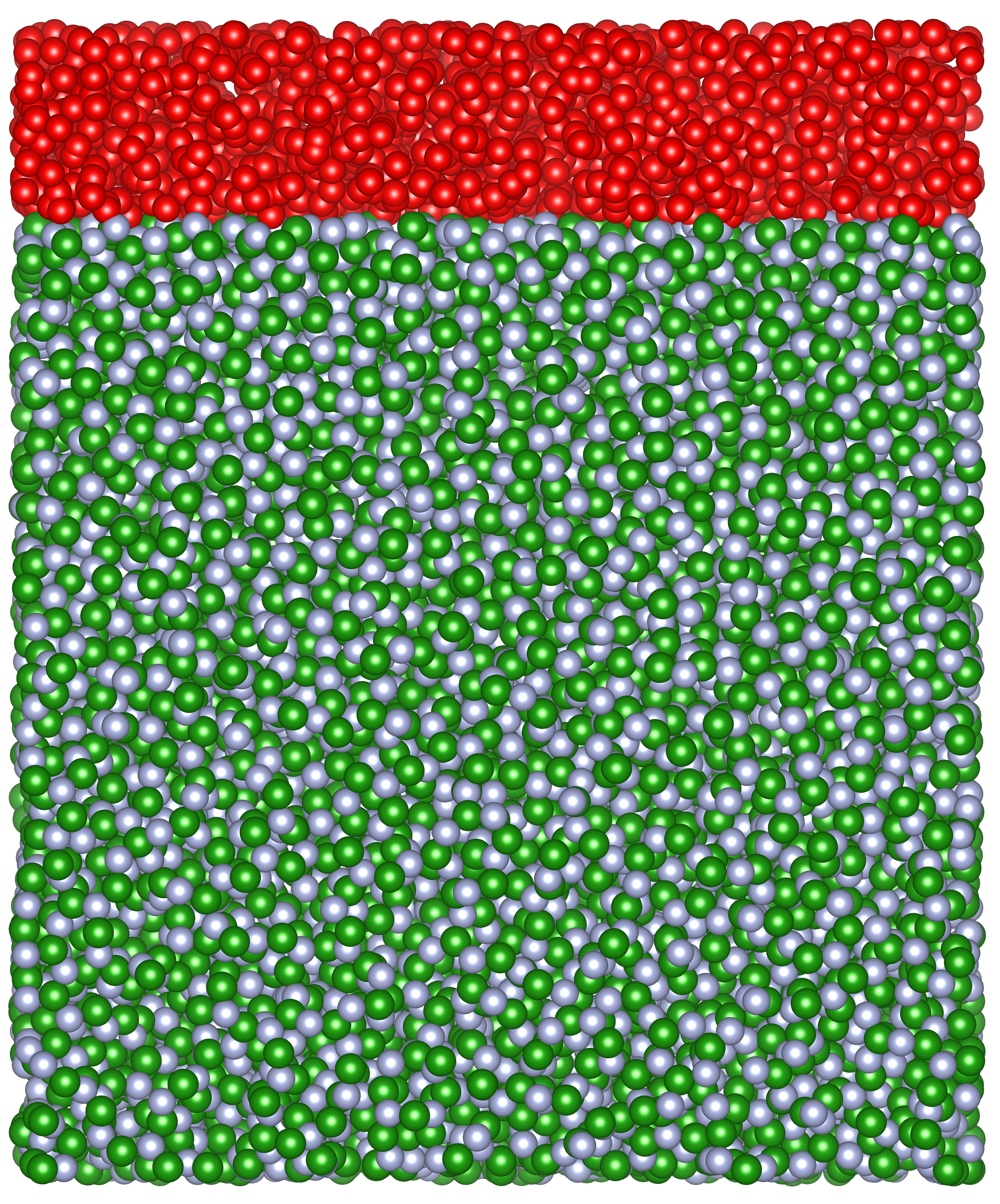}
\caption{Simulation cell used to model oxidation of \textrm{$\alpha$}-BN. B, N, and O atoms are shown in green, white, and red, respectively. The 5 nm-thick \textrm{$\alpha$}-BN film is supported by a frozen BN layer (omitted for clarity), and a 5 nm vacuum region above the film contains randomly distributed atomic oxygen to simulate an oxidative environment. Visualized using VESTA \cite{vesta}.}
\label{fig:unitcell}
\end{figure}

\begin{figure}[htbp]
  \centering
  \begin{subfigure}[t]{0.45\textwidth}
    \centering
    \includegraphics[width=\linewidth]{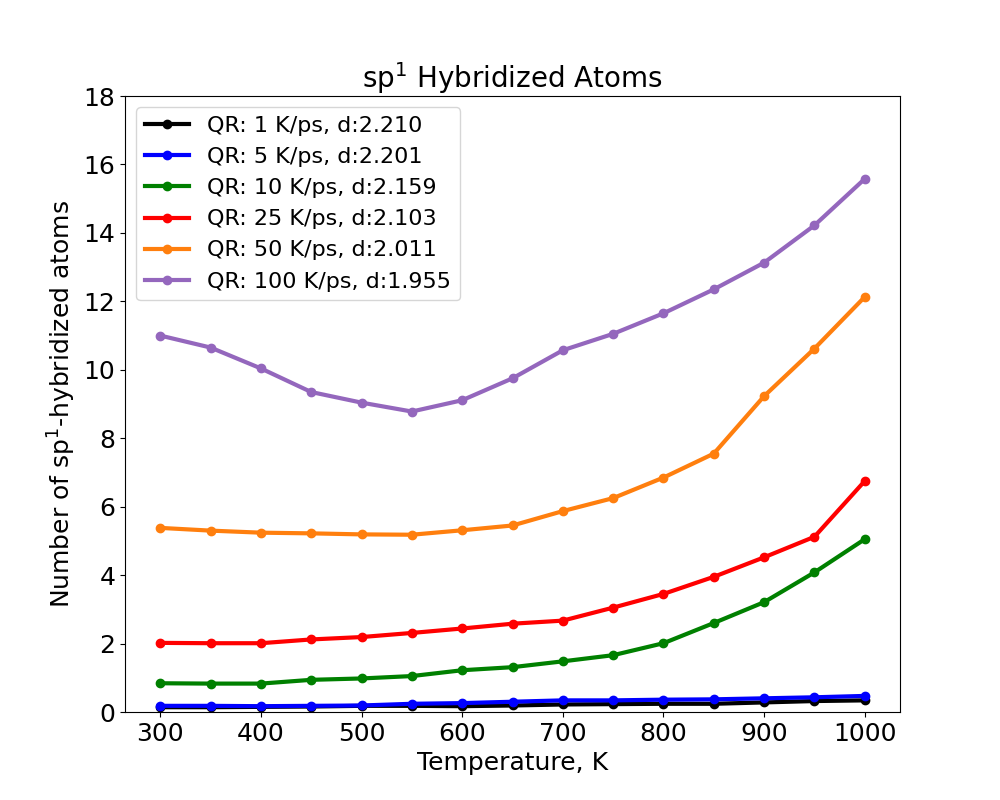}
    \caption{}
    \label{fig:sp1_oxid} 
  \end{subfigure}
  \hspace{0.05\textwidth}
  \begin{subfigure}[t]{0.45\textwidth}
    \centering
    \includegraphics[width=\linewidth]{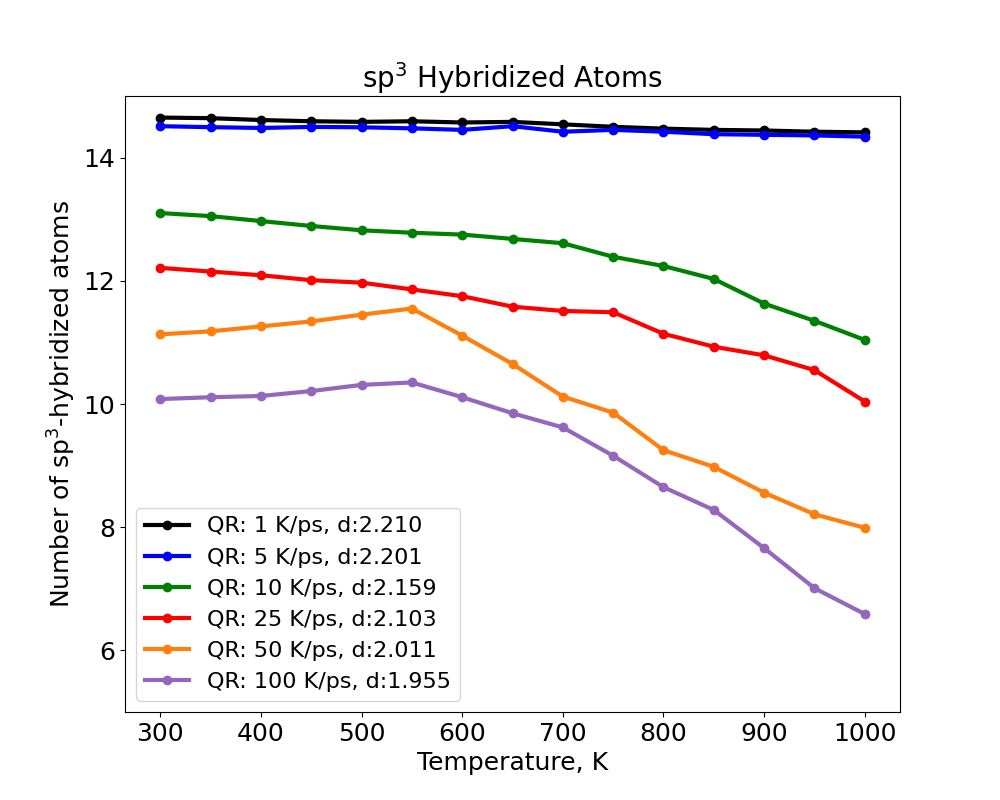}
    \caption{}
    \label{fig:sp2_oxid} 
  \end{subfigure}
  \label{fig:hybrid_oxid}
  \caption{Evaluation of hybridization states during oxidation for \textrm{$\alpha$}-BN films prepared at different quench rates. (a) \textrm{$sp$} fraction. (b) \textrm{$sp^2$} and \textrm{$sp^3$} fractions. }
\end{figure}

We examine how film morphology governs oxidation of \textrm{$\alpha$}-BN using MD with machine-learned GAP potentials, as described in the Methods. Films were generated by varying the quenching rate from 1 to 100 K ps$^{-1}$, spanning chemically ordered, high-density continuous random networks to porous, defect-rich structures, which enables a direct comparison of oxidation across morphologies. The simulation cell contains atomic oxygen occupying 20\% of the vacuum region to raise the oxygen chemical potential as shown in Figure \ref{fig:unitcell}. All films follow the same stepwise heating with equilibration at each stage to 1000 K. \\
We therefore focus on mechanisms and depth trends rather than absolute rates. The results reveal a strong structure–reactivity relation. Slowly quenched films, with minimal homonuclear bonding, near-stoichiometric composition, and predominantly \textrm{$sp^2$} hybridization, resist oxygen incorporation even at high temperature and retain structural integrity. Fast quenched films, with more under-coordinated sites, higher B–B and N–N fractions, and lower mass density, oxidize at lower temperatures and exhibit deeper oxygen penetration with associated structural degradation. These observations indicate that oxidation resistance in \textrm{$\alpha$}-BN is controlled by local structural features, including under-coordinated atoms, homonuclear clusters, and nanoporosity. The remainder of this section analyzes these processes in detail, tracking changes in bonding, hybridization, and stoichiometry.\\
\begin{figure}[htbp]
  \centering
  \begin{subfigure}[t]{0.45\textwidth}
    \centering
    \includegraphics[width=\linewidth]{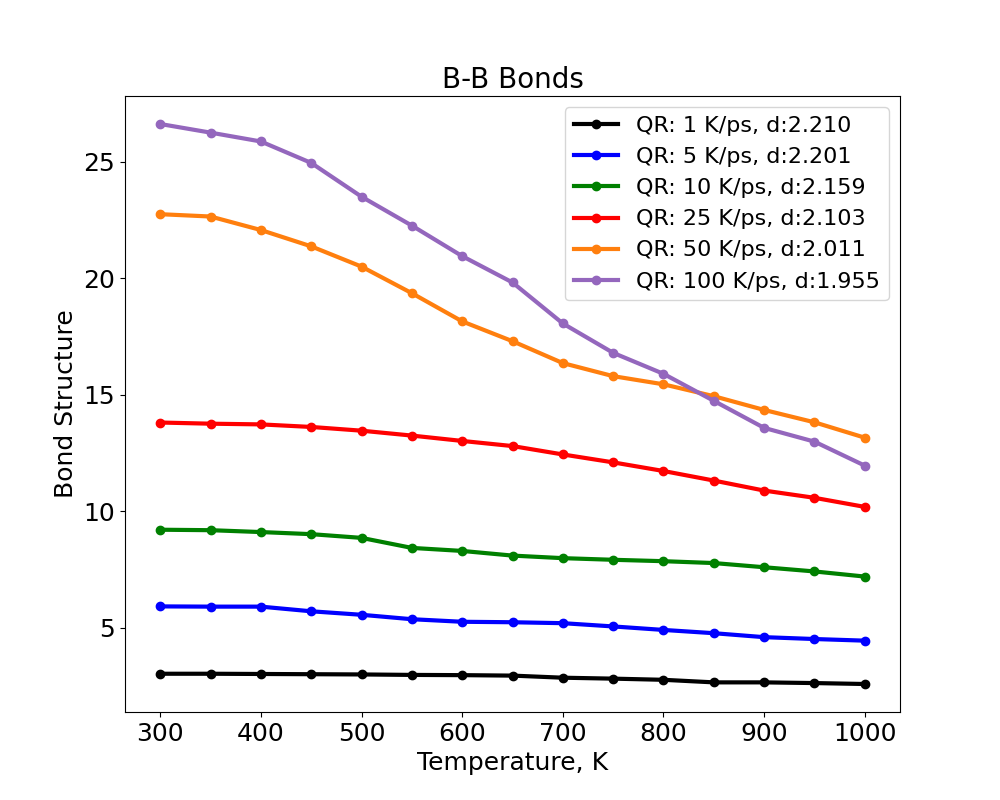}
    \caption{}
    \label{fig:B_B_oxid} 
  \end{subfigure}
  \hspace{0.05\textwidth}
  \begin{subfigure}[t]{0.45\textwidth}
    \centering
    \includegraphics[width=\linewidth]{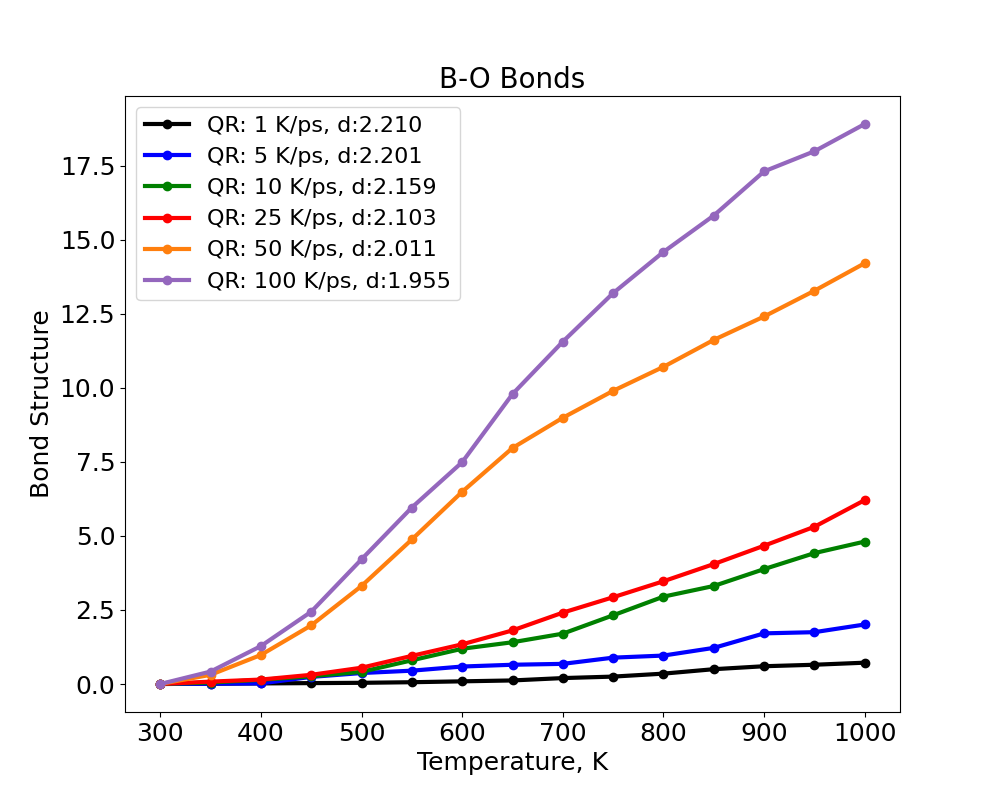}
    \caption{}
    \label{fig:B_O_oxid} 
  \end{subfigure}
  \vspace{1em}
  \begin{subfigure}[t]{0.45\textwidth}
    \centering
    \includegraphics[width=\linewidth]{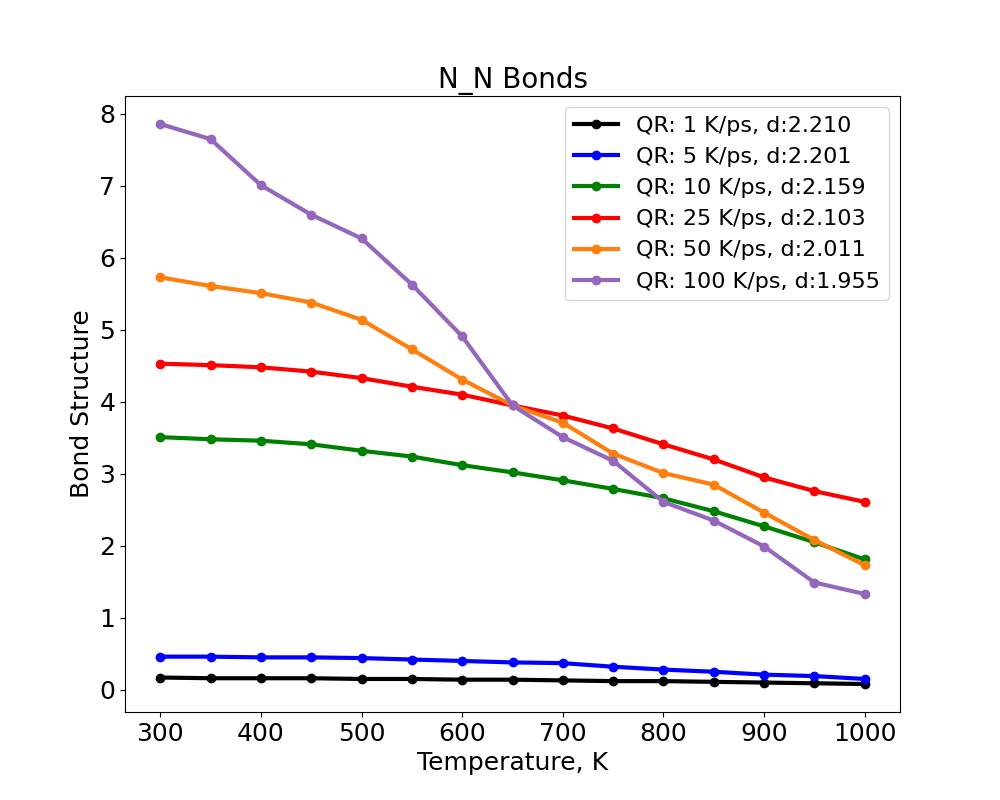}
    \caption{}
    \label{fig:N_N_oxid} 
  \end{subfigure}
  \hspace{0.05\textwidth}
  \begin{subfigure}[t]{0.45\textwidth}
    \centering
    \includegraphics[width=\linewidth]{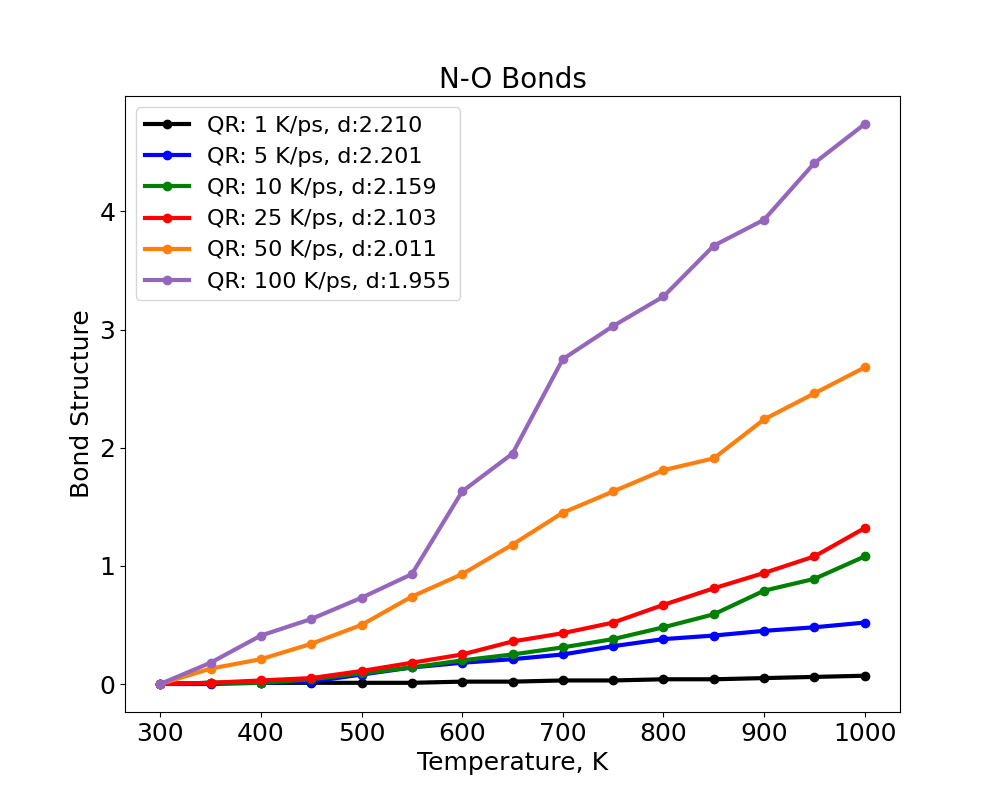}
    \caption{}
    \label{fig:N_O_oxid} 
  \end{subfigure}
  \caption{(a) Fraction of B–B bonds, (b) B–O bonds, (c) N–N bonds, and (d) N–O bonds as a function of temperature during oxidation of \textrm{$\alpha$}-BN films generated at different quench rates.  }
  \label{fig:bond_hybrid}
\end{figure}
The morphology of films after generation determines their oxidation mechanism. Figure \ref{fig:bond_hybrid} shows distinct differences between the behavior of the films. slowly quenched (1-5 K/ps) films having high density, less free space, and mostly B-N bonds ($>$ 90\%) exhibit remarkable chemical stability under heating conditions and oxidizing environment. The concentration of B-O and N-O bonds remains negligible across the entire temperature range, increasing by less than 2\% even at 1000 K (Figure \ref{fig:B_O_oxid}, \ref{fig:N_O_oxid}). The small number of pre-existing B–B and N–N defect bonds in these films actually decreases slightly upon oxidation, suggesting that oxygen preferentially reacts at these rare defect sites. Once these vulnerable sites are oxidized, there are no accessible pathways for further oxygen infiltration or attack on the well-connected B–N network. Therefore, in these films, we only observed oxidation at surface levels. However, with larger quenching rates, initial morphologies have more homonuclear bonds as discussed in Section 2.1. Fast quenched films (50 - 100 K/ps) have $>$20\% B-B bonds and $>$5\% N-N bonds. Considering higher \textrm{$sp$}-hybridized atoms (Figure \ref{fig:sp1_oxid}), these films are less stable and may have more free volume that might allow O atoms to penetrate into the films. With the introduction of the heat and oxygen, these films quickly start to lose B-B and N-N atoms. Subsequently, B-O and N-O bonds are also increased. B–N bonds remain essentially unchanged, with only a small decrease, whereas homonuclear B–B and N–N bonds oxidize rapidly. Formation of B–O and N–O bonds is dominated by oxidation at homonuclear B–B and N–N sites. A minor fraction arises from oxidation of \textrm{$sp$} and \textrm{$sp^2$} B and N within the B–N network. Some N–N bonds break, and nitrogen leaves as \textrm{$N_2$} or \textrm{$NO_x$}. The resulting vacancies open pathways for oxygen to reach the interior, promoting oxidation of homonuclear bonds at depth. Therefore, in these films, we observe a bulk oxidation rather than a surface level one as in slow-quenched films. Intermediate quench rates result in behavior between these two limits, with partial oxidation confined near the surface rather than progressing into the bulk, as seen in fast-quenched films, while exhibiting less resistance than the most slowly quenched samples. These films exhibit moderate disorder and partial oxidation. While B-B bonds are reduced slightly, a steady increase in B-O bonds reaches 4.81\% and 6.22\% in films generated at 10 K/ps and 25 K/ps, respectively. The reduction in N–N is more pronounced, while N–O rises only slightly above 1\%, suggesting nitrogen more often leaves the structure than remains as N–O. This progression, starting with selective oxidation at pre-existing defects and advancing toward network-wide degradation enabled by nitrogen loss and free volume, defines the shift from surface-limited to bulk oxidation across quench rates.\\
\begin{figure}
    \centering
    \includegraphics[width=0.7\linewidth]{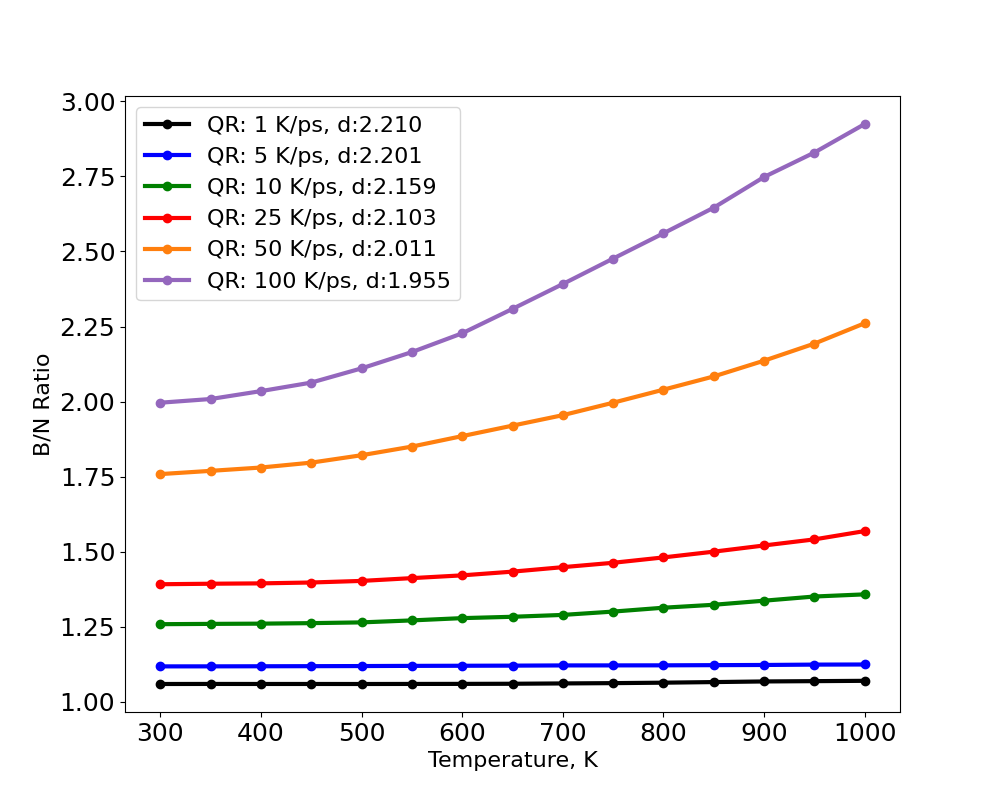}
    \caption{B/N atomic ratio versus temperature during oxidation for \textrm{$\alpha$}-BN films prepared at different quench rates.}
    \label{fig::B_N_rat_oxid}
\end{figure}
In addition to changes in bonding environments, variations in atomic hybridization provide critical insight into the structural response under oxidation. As discussed in Section 2.1, applied quench rates result in a diverse set of atomic hybridization states. Slowly quenched films contain mostly \textrm{$sp^2$} atoms ($>$85\%), with the remainder primarily \textrm{$sp^3$}, except for a few under-coordinated sites near the surface. This results in a stable network that remains unchanged under heating and oxygen exposure, as discussed above. As a result, the hybridization state remains virtually unchanged throughout the simulation. By contrast, structures formed at intermediate and fast QR show changes with increasing temperature. This is attributed due to their less stable and more reactive network. Upon oxygen exposure and heating, \textrm{$sp$} atoms—primarily dangling bonds near the surface—are passivated, leading to a decrease in \textrm{$sp$} and a corresponding increase in \textrm{$sp^2$} atoms. A similar but less apparent increase is observed in the 50 K/ps sample.However, with increasing temperatures, we observe an increase in \textrm{$sp$}-hybridized atoms. Nitrogen loss at elevated temperature leaves neighboring atoms under-coordinated. Additionally, oxygen incorporation into deeper regions introduces local strain, increasing porosity and free volume. These factors accumulate dangling bonds and promote defect formation. These changes are less evident in intermediate QR films as shown in Figure \ref{fig:hybridmorp}. At low temperatures, the \textrm{$sp$} fraction remains nearly constant: oxygen passivates coordination defects, but nitrogen loss and strain from B–O and N–O formation introduce new under-coordinated sites. Increased \textrm{$sp$}-hybridized atoms, together with the changing bonding environment, shows how much the atomic environment changes during the heating cycle. These hybridization changes reveal how initial coordination disorder enables deeper oxygen infiltration by destabilizing the network and regenerating reactive sites during oxidation.\\
The B/N ratio provides a quantitative measure of chemical evolution during oxidation. Figure \ref{fig::B_N_rat_oxid} shows distinct trends in the evolution of the B/N ratio. As discussed above, dense and highly stable \textrm{$\alpha$}-BN films show minimal change under heating and oxygen exposure. Because oxidation is confined to the surface and the films contain few homonuclear bonds, the B/N ratio remains constant and near-stoichiometric. In contrast, intermediate quench rates result in more N–N bonding and an initially elevated B/N ratio. Additionally, oxygen atoms oxidize these nitrogen-rich regions, and nitrogen is lost in the form of \textrm($NO_2$). These structural changes are even more apparent in films quenched at high rates (50–100 K/ps).The temperature-driven rise in the B/N ratio confirms that nitrogen depletion—initiated by homonuclear bond breakage and oxygen reactions—plays a central role in enabling bulk oxidation in disordered films.\\
\begin{figure}[ht]
    \centering
    \begin{subfigure}[b]{0.32\textwidth}
        \centering
        \includegraphics[width=\textwidth]{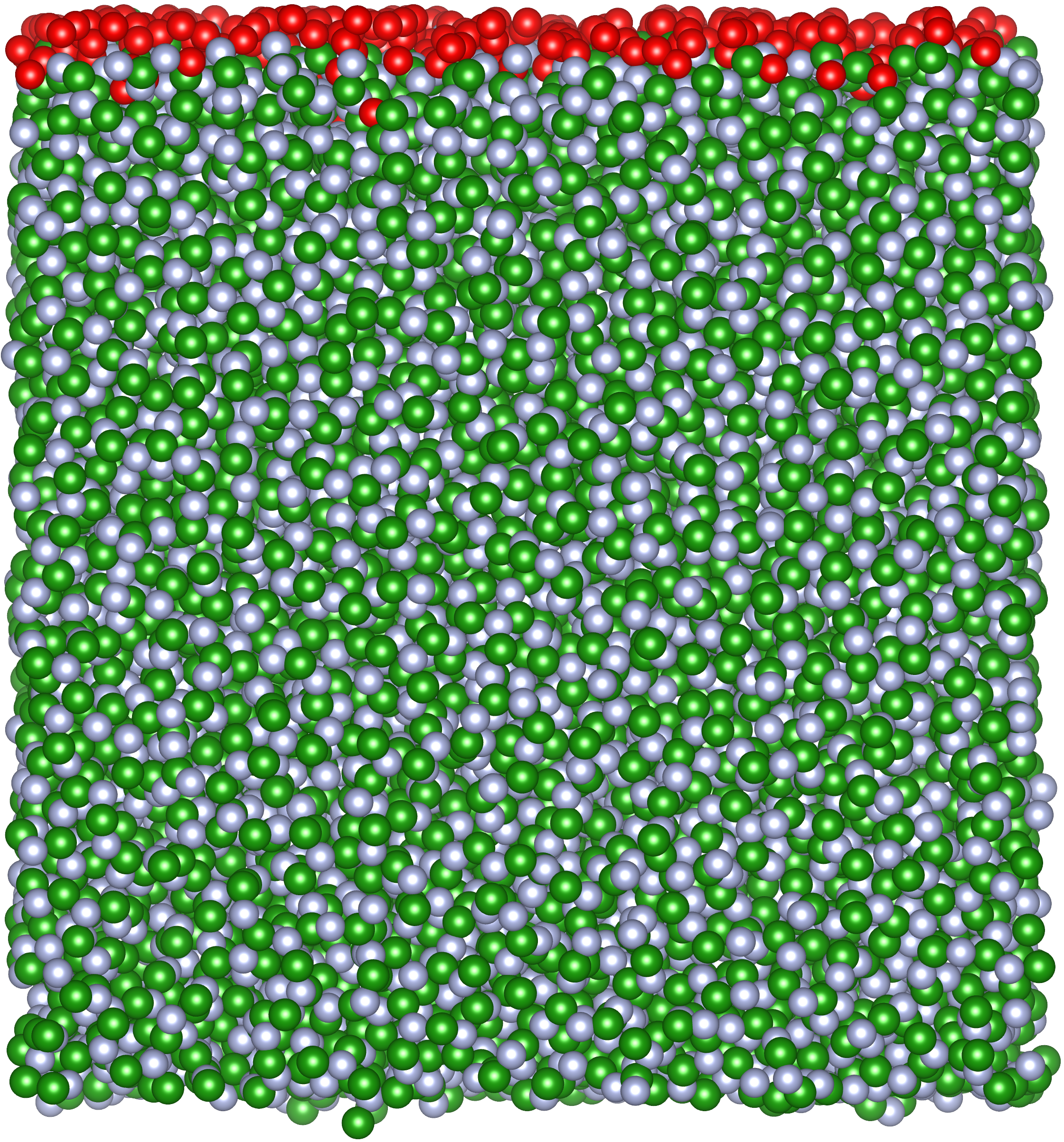}
        \caption{ }
        \label{fig:1kps}
    \end{subfigure}
    \hfill
    \begin{subfigure}[b]{0.32\textwidth}
        \centering
        \includegraphics[width=\textwidth]{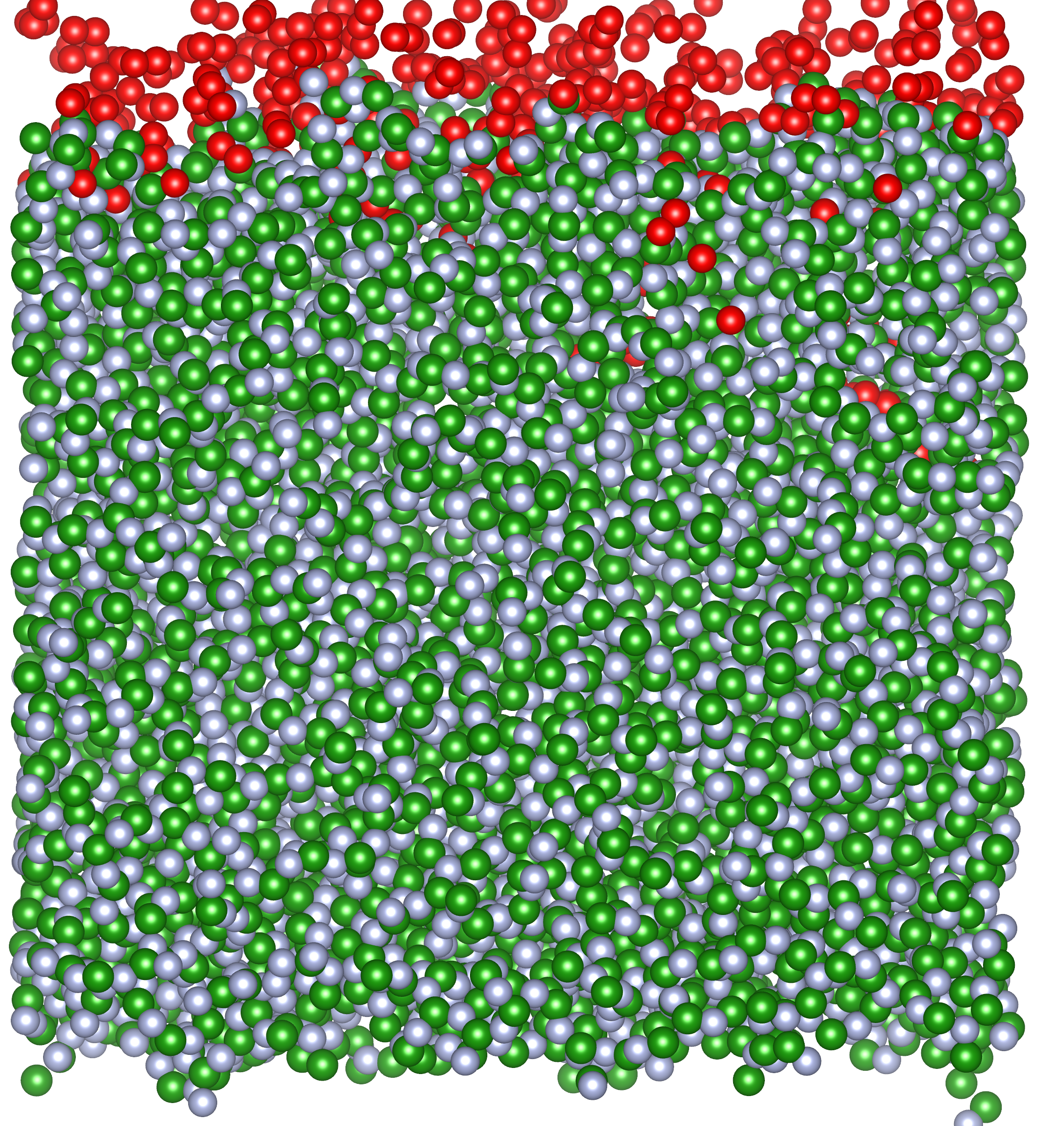}
        \caption{ }
        \label{fig:25kps}
    \end{subfigure}
    \hfill
    \begin{subfigure}[b]{0.32\textwidth}
        \centering
        \includegraphics[width=\textwidth]{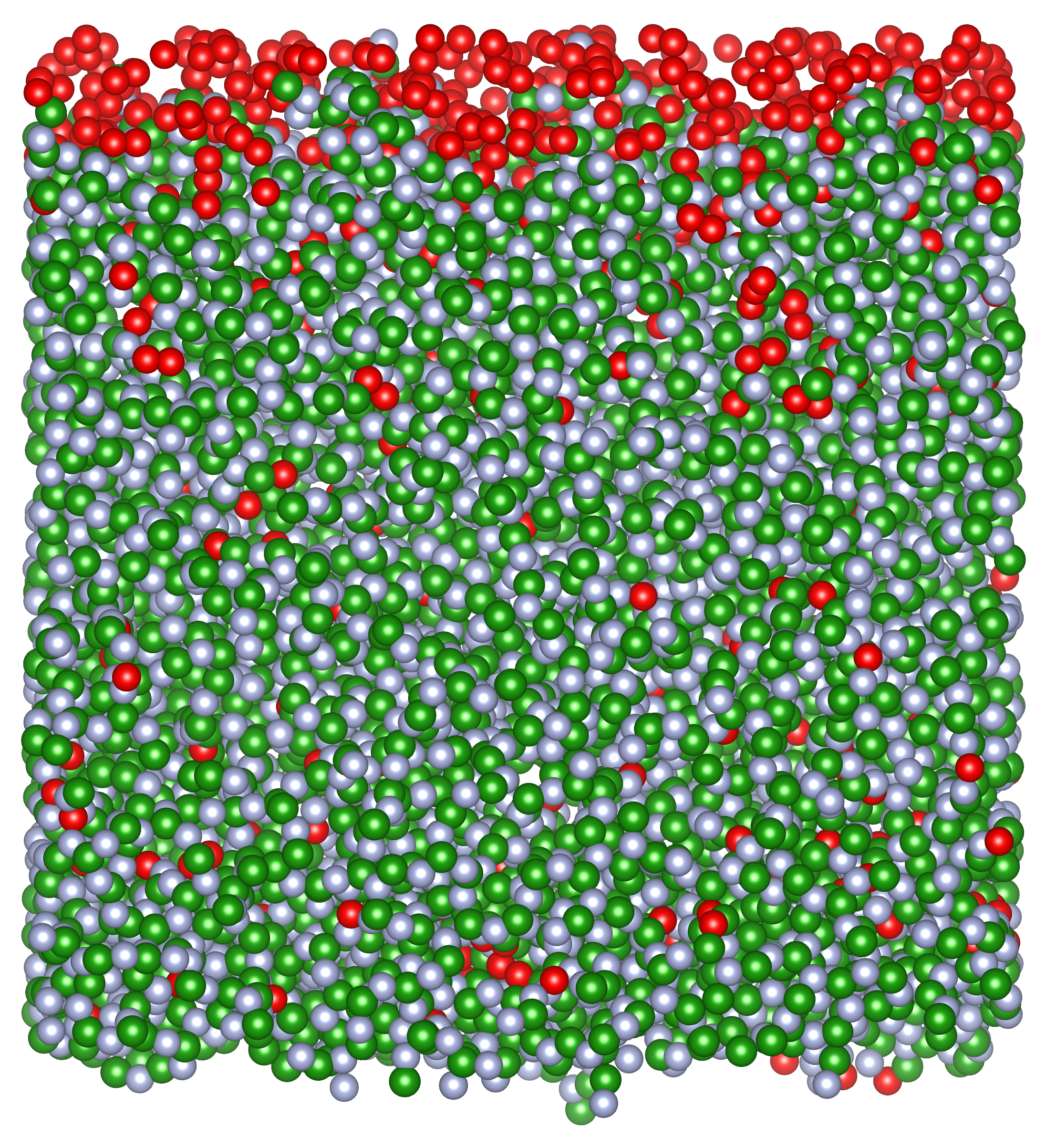}
        \caption{ }
        \label{fig:100kps}
    \end{subfigure}
    \caption{Snapshots of \textrm{$\alpha$}-BN films generated at 1 $  K,ps^{-1}$, 25 $  K,ps^{-1}$, and 100 $  K,ps^{-1}$ after oxidation at 1000 K. Oxygen atoms are initially placed above the \textrm{$\alpha$}-BN layer and penetrate to different depths depending on the film morphology. Atoms above the film are partially removed in the visualisation to improve clarity. Visualized using VESTA \cite{vesta}.}
    \label{fig:oxidation_snapshots}
\end{figure}
These simulations establish a coherent framework connecting the structural characteristics of amorphous BN films to their oxidation behavior. Figure~\ref{fig:oxidation_snapshots} shows final atomic configurations under identical oxygen loading and thermal treatment. Oxygen remains near the surface in the dense 1 K/ps film, is partly incorporated at 25 K/ps, and extends below the surface at 100 K/ps, matching the disorder gradient quantified below. Films formed at different quench rates develop distinct local bonding, hybridization, and defect landscapes that dictate oxygen accessibility and reaction progression. Surface-limited oxidation in dense, more ordered networks arises from scarce homonuclear bonds and few under-coordinated sites. Disordered films, rich in B–B and N–N bonds, \textrm{$sp$}-hybridized atoms, and free volume, undergo nitrogen depletion, internal bond scission, and oxygen penetration that drive bulk oxidation. Intermediate structures display mixed behavior. Partial oxidation is enabled by moderate reactivity but constrained by residual connectivity. The evolution of bonding, hybridization, and the B/N ratio across this disorder spectrum defines a mechanistic pathway for oxidation that is structurally driven and chemically distinct. These findings clarify the role of atomic-scale disorder in \textrm{$\alpha$}-BN degradation and identify measurable signatures of oxidation resistance.
\subsection{XPS Analysis} \label{s:experiment}
\begin{figure}[htbp]
\centering
\includegraphics[width=\columnwidth]{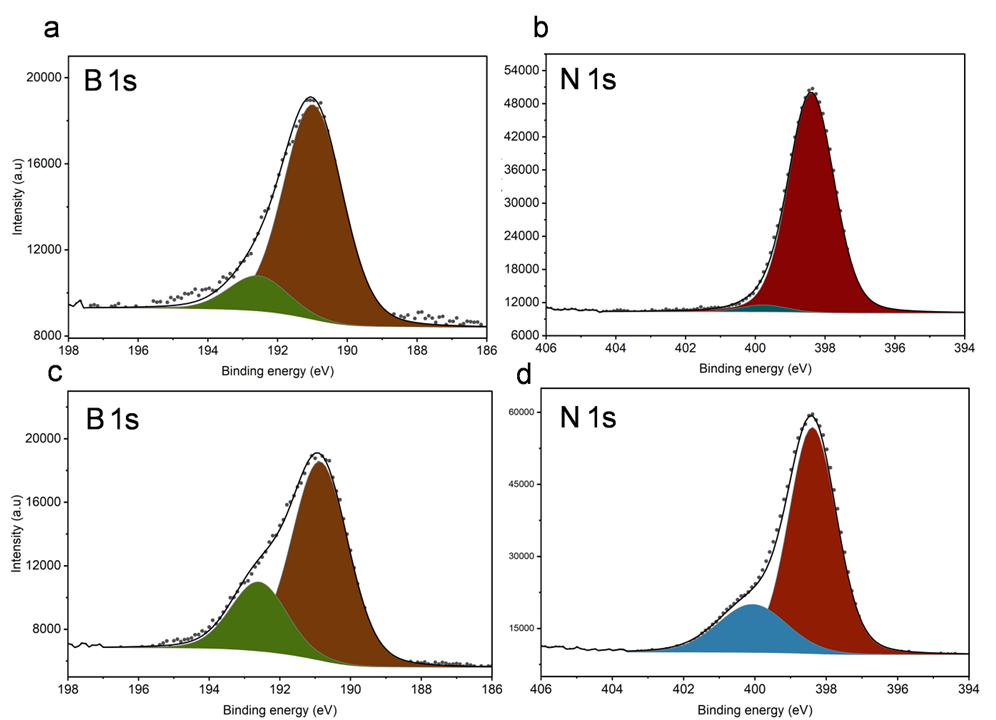}
\caption{XPS core-level spectra of sample surface collected at 90° take-off angle before air exposure. (a) B 1s spectrum of \textrm{$\alpha$}-BN-900; (b) N 1s spectrum of \textrm{$\alpha$}-BN-900; (c) B 1s spectrum of \textrm{$\alpha$}-BN-800; (d) N 1s spectrum of \textrm{$\alpha$}-BN-800.}
\label{fig:XPS_before}
\end{figure}

To experimentally validate the oxidation mechanisms proposed by our simulations, we investigated the chemical stability of \textrm{$\alpha$}-BN films grown at 900 °C (\textrm{$\alpha$}-BN-900) and 800 °C (\textrm{$\alpha$}-BN-800) as representative cases using XPS. Both films were evenly deposited onto a silicon wafer using CVD. The simulation results suggests that growth temperature governs local bonding features and defect content, consequently influencing the susceptibility of under-coordinated boron sites to oxidation. The experimental challenge lies in determining whether these structural differences translate into measurable changes in environmental reactivity. By capturing the depth-dependent evolution of bonding environments before and after ambient air exposure, the XPS measurements are expected to evaluate the occurrence and extent of oxidative modification in both films.\\
The XPS spectra of \textrm{$\alpha$}-BN-800 and \textrm{$\alpha$}-BN-900 before exposure are presented in Figure \ref{fig:XPS_before}. The decomposed sub-peaks with the highest intensity centred at 190.7 and 398.4 eV for B 1s and N 1s, respectively, can be attributed to B-N binding. For B 1s, no signal is detected at 187.3 eV, where B-B bonding is typically assigned, indicating that boron atoms are predominantly coordinated with nitrogen. It is worth noting that the spectrum lacks \textrm{$sp$}-related satellite features, and minimal changes are observed with varying take-off angles, signifying a disordered bonding network with less crystallinity (\ref{fig:XPS_before}a,c and \ref{supp-fig:XPS_core_B}a,c). In addition, the sub-peak of B 1s at 192.5 eV corresponds to BN oxidation, presumably \textrm{$B_2O_3$}. In contrast, the lack of direct N-O binding at high binding energy of N 1s can be ascribed to the chemical inertness of nitrogen sites and a robust B-N framework (\ref{fig:XPS_before}-b and \ref{fig:XPS_before}-d). Nonetheless, the sub-peak of N 1s centred at 400-400.7 eV is noted for both cases, possibly related to the presence of N-H binding. The low volume of this remaining N-H bond also contributes to the framework stability of \textrm{$\alpha$}-BN-900. Note that the relatively low sub-intensity ratio in \textrm{$\alpha$}-BN-900 (1:6) than that of \textrm{$\alpha$}-BN-800 (1:3) also suggests that \textrm{$\alpha$}-BN-900 is less oxygen sensitive. Note also that, in both cases, the increasing intensity with decreased take-off angle indicates that oxidation takes place mostly at the topmost surface of the film (Figure \ref{supp-fig:XPS_core_B} and \ref{supp-fig:XPS_core_N}).\\

\begin{figure}[htbp]
\centering
\includegraphics[width=\columnwidth]{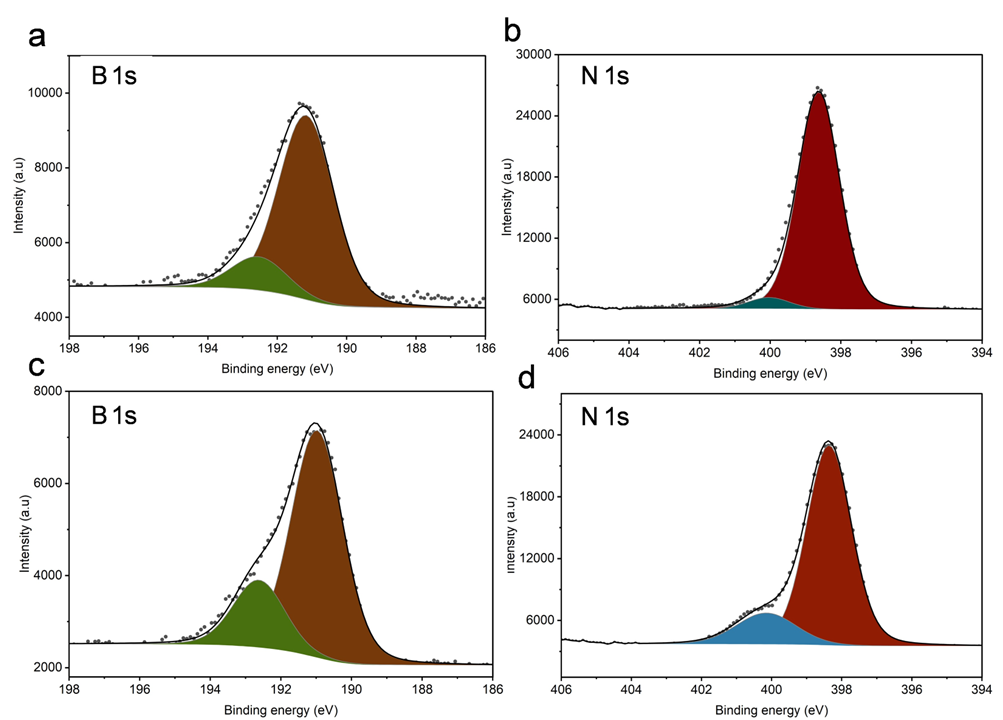}
\caption{XPS core-level spectra of sample surface collected at 90° take-off angle after air exposure. (a) B 1s spectrum of \textrm{$\alpha$}-BN-900; (b) N 1s spectrum of \textrm{$\alpha$}-BN-900; (c) B 1s spectrum of \textrm{$\alpha$}-BN-800; (d) N 1s spectrum of \textrm{$\alpha$}-BN-800.}
\label{fig:XPS_after}
\end{figure}
Upon exposure, the B-O signal shows a slight increase for both cases (Figure \ref{fig:XPS_after} and Table \ref{supp-tab:angles}), whereas the main amorphous B-N peak remains unchanged in both position and shape. These results indicate that oxidation is confined to the near-surface region, without significant alteration of the bulk bonding structure. In particular, \textrm{$\alpha$}-BN-900 demonstrates sound oxidation resistance, suggested by insignificant increase of B-O peak sub-intensity ratio (1:5) after exposure. XPS depth profile also reveals that oxidation mainly occurs at the outer surface, evidenced by the  increase in intensity of B-O at decreased take-off angles (Figures \ref{supp-fig:XPS_core_B}, \ref{supp-fig:XPS_core_N} and Table \ref{supp-tab:angles}). In addition, the N 1s profile demonstrates insignificant change in chemical environment, i.e., nitrogen oxidation, before and after exposure, suggesting chemical relative inertness of nitrogen sites and a robust B-N framework (Figure \ref{supp-fig:XPS_core_N}). Together, these observations point to \textrm{$\alpha$}-BN-900 as a chemically stable amorphous film with more pronounced oxidation resistance, in agreement with the more ordered, nitrogen-rich structure formed at elevated growth temperature. \\
These spectroscopic observations are consistent with the oxidation mechanisms predicted by our simulation. In both films, the evolution of the B 1s spectrum after air exposure is confined to a modest increase in the B-O component without any major shifts in the B-N peak position or line shape, indicating that oxygen incorporation is largely restricted to the surface or near-surface region. The minimal angular dependence and absence of major changes in the chemical environment confirm that oxygen does not penetrate the bulk of the films. This behavior shows good agreement with the simulation results on the intermediate-quenching-rate structures (10–25 K/ps), where oxygen preferentially reacts with under-coordinated boron atoms and pre-existing structural defects rather than diffusing through the network. The partial oxidation observed at the surface of \textrm{$\alpha$}-BN-800, and to a lesser extent in \textrm{$\alpha$}-900, can therefore be attributed to defect-mediated processes, whereby oxygen atoms are accommodated at local coordination vacancies or disrupted regions within the amorphous network. The preservation of the N 1s peak shape and binding energy further confirms the stability of nitrogen environments and rules out substantial nitrogen oxidation or network degradation. Together, these results demonstrate that oxidation proceeds through localised reactions at defect sites, while the bulk B-N framework remains chemically inert. This validates the theoretical picture of a defect-limited oxidation mechanism and highlights the critical role of structural ordering and coordination quality in determining the environmental stability of amorphous BN films.
\section{Conclusions}
This study provides atomistic insight into the oxidation behavior of \textrm{$\alpha$}-BN by combining large-scale machine learning molecular dynamics and angle-resolved XPS experiments. We show that the chemical stability of \textrm{$\alpha$}-BN is not an intrinsic property but is instead governed by its complex morphology, bonding environment, and defect concentration, which are in turn shaped by synthesis conditions. The simulations reveal that oxygen preferentially attacks under-coordinated boron sites and nitrogen-deficient regions, triggering bond rearrangements and gradual network degradation, particularly in low-density, porous films with high fraction of $  sp$-hybridized atoms. In contrast, ordered, nitrogen-rich networks formed under slower cooling rates resist oxidation and preserve B–N network. These mechanistic trends are consistent with experimental XPS data on two \textrm{$\alpha$}-BN films synthesized at different temperatures by CVD, which show minimal shifts in the B 1s and N 1s spectra after prolonged air exposure, indicating that oxidation remains confined to surface-proximal regions and does not disrupt the bulk structure.The film synthesized at a higher temperature exhibits greater resistance to oxidation, a property attributed to its more ordered structure which features a B/N ratio nearer to stoichiometric and consequently fewer sites for oxygen to attack. In contrast, the lower-temperature analogue's enhanced oxygen reactivity is consistent with a higher initial concentration of structural defects. These findings collectively demonstrate that the oxidation resistance of amorphous BN can be precisely tuned through synthesis parameters that govern its morphology and bonding, offering a clear pathway for designing chemically robust, ultrathin diffusion barriers for advanced nanoelectronic applications.
\section{Methods} \label{s:methods}
\subsection{Molecular Dynamic Simulations} \label{s:MD}
All molecular dynamics (MD) simulations were performed using the LAMMPS package and a machine learning-based Gaussian Approximation Potential (GAP) model developed specifically for B–N–O systems. Details of the dataset construction and GAP training procedure are provided in the Supplementary Information.\\
\textrm{$\alpha$}-BN thin films were generated via a melt–quench protocol. The initial simulation cell contained a random distribution of B and N atoms within a 50 $ \textrm{\AA}$ × 50 $ \textrm{\AA}$ × 100 $ \textrm{\AA}$ box, with 25 $ \textrm{\AA}$ of vacuum added in both the $ +z $ and $ -z $ directions to model freestanding thin films. Following energy minimization, each system was equilibrated at 5000 K in the NVT ensemble (constant volume and temperature) for 20 ps using a 0.25 fs timestep. The structures were then cooled to 300 K at quenching rates between 1 and 100 K ps$^{-1}$, followed by NVE (constant energy) relaxation and additional NVT equilibration for 20 ps each. To match experimental conditions, the initial atomic density was fixed at approximately 2.1 g/cm$^3$. Unbonded atoms were removed from the final structures to ensure stoichiometrically consistent \textrm{$\alpha$}-BN networks.\\
The resulting films varied in density, bonding topology, and hybridization, depending on the cooling rate. Slowly quenched films formed dense, chemically ordered networks with few coordination defects, while fast-cooled samples showed increased porosity, a higher fraction of homonuclear bonds, and under-coordinated sites. For oxidation studies, a separate 1 nm-thick layer of pre-relaxed BN was inserted below the amorphous films and kept frozen throughout the simulations to represent a passive substrate and to suppress system-wide relaxation artifacts. Atomic oxygen was introduced into the upper vacuum region, randomly distributed and free of initial overlaps as shown in Figure \ref{fig:unitcell}. Oxidation was simulated through a series of heating-equilibration cycles: the temperature was increased in 100 K increments, each followed by 50 ps of heating and 50 ps of equilibration under the NVT ensemble, up to a final temperature of 1000 K. Atomic trajectories, bonding configurations, hybridization states, and stoichiometry were tracked at each stage to evaluate how initial film morphology affects oxidation mechanisms. This approach enabled a systematic investigation into the structure–oxidation relationship across a spectrum of disordered \textrm{$\alpha$}-BN films.\\
Using atomic O intentionally increases the oxygen chemical potential and compresses time scales, enabling access to the same mechanistic channels expected under air without claiming kinetic fidelity. We therefore compare mechanisms and depth trends rather than absolute rates \cite{Li2024machine, Cvitkovich2024}.

\subsection{Sample Synthesis}
Thin BN layers were deposited at two different temperatures on Si (100) with native oxide by chemical vapour deposition. Depositions took place at either 800 °C or 900 °C in a home-made reactor composed of a tubular horizontal furnace as reaction chamber with a base pressure of 1.4 mbar. The substrates are raised in deposition temperature with 10 °C.min$^{-1}$ ramp under 5\% hydrogenated argon flux. Contained at 0 °C in a stainless steel canister, borazine was introduced in the reaction chamber as unique B and N source. Deposition was performed for 10 min under 100 sccm of 5\% hydrogenated argon. The reactor base pressure being 100 fold lower than the vapour pressure of borazine at 0 °C (100 mbar), flash evaporation occurred. The film thickness was then adjusted as a function of the reaction temperature and borazine amount.\\

\subsection{Sample characterization}

The thickness of thin films was determined by spectroscopic ellipsometry at an incident angle of 75° using a Semilab (SE-2000) ellipsometer. Data fit was realized with the ‘SAM suite’ software using a Cauchy dispersion law. Raman spectroscopy was also performed using a LabRAM HR Evolution (Horiba) spectrometer with a 632 nm excitation wavelength.\\
XPS measurements were performed using a Kratos AXIS Supra+ spectrometer equipped with a monochromated Al K\textrm{$\alpha$} X-ray source (h$  \nu$ = 1486.7 eV). The base pressure in the analysis chamber was approximately 5 × 10$^{-8}$ mbar. Instrument work function calibration was conducted using the Cu LMM Auger peak. A coaxial charge neutraliser was applied during all measurements to minimise surface charging artifacts.\\
Survey spectra were acquired over a 0.7 × 0.4 mm$^2$ analysis area using a pass energy of 160 eV and a step size of 1 eV. High-resolution spectra were collected from the same area with a pass energy of 40 eV and a step size of 0.1 eV. The X-ray source operated at 90 W, and the photoelectron take-off angle was set at 90°.\\
Angle-resolved XPS (ARXPS) was carried out using the same instrument at take-off angles of 20°, 27°, 35°, 45°, 50°, and 90° relative to the sample surface. With increasing take-off angle, the effective analysis depth increased, reaching up to approximately 10 nm. Lower angles provided information on the outermost surface layers, whereas higher angles yielded insight into subsurface regions approaching the bulk.\\
All spectra were processed using CasaXPS software (version 2.3.24 PR1.0, Casa Software Ltd). Shirley background subtraction was applied to all peaks. A generalized Lorentzian line shape LA(1.53, 243), as implemented in the software, was used for peak fitting. Charge correction was performed by referencing the C 1s peak (C–C bond) to 285.0 eV.

\begin{acknowledgement}

This project is conducted under the REDI Program, a project that has received funding from the European Union's Horizon 2020 research and innovation programme under the Marie Skłodowska-Curie grant agreement no. 101034328. This paper reflects only the author's view and the Research Executive Agency is not responsible for any use that may be made of the information it contains.S.R acknowledge funding from European Union “NextGenerationEU/PRTR” under grant PCI2021-122035-2A-2a. ICN2 is funded by the CERCA Programme/Generalitat de Catalunya and supported by the Severo Ochoa Centres of Excellence programme, Grant CEX2021-001214-S, funded by MCIN/AEI/10.13039.501100011033. This work is also supported by MICIN with European funds- NextGenerationEU (PRTR-C17.I1) and by 2021 SGR 00997, funded by Generalitat de Catalunya. Simulations were performed at the King Abdullah University of Science and Technology-KAUST (Supercomputer Shaheen II Cray XC40) and the Center for Nanoscale Materials, a U.S. Department of Energy Office of Science User Facility, supported by the U.S. DOE, Office of Basic Energy Sciences, under Contract No. DE-AC02-06CH11357.

\end{acknowledgement}

\begin{suppinfo}

The following file is available free of charge. \\ SupportingInformation.pdf: Methods and additional results for the machine-learned interatomic potential, including training-set composition, GAP hyperparameters, and validation plots; supplementary characterization (Raman spectra) of CVD-grown films; full XPS spectra before and after ambient exposure at multiple take-off angles; and tabulated compositions and angle-resolved ratios used in the main text.

\end{suppinfo}

\bibliography{achemso-demo}

\end{document}